\documentclass[fleqn,usenatbib]{mnras}

\usepackage[T1]{fontenc}

\DeclareRobustCommand{\VAN}[3]{#2}
\let\VANthebibliography\thebibliography
\def\thebibliography{\DeclareRobustCommand{\VAN}[3]{##3}\VANthebibliography}

\usepackage{graphicx}	
\usepackage{amsmath}	
\usepackage{amssymb}	

\title[An updated catalogue of QHCs candidates]{The population of Comet candidates among quasi-Hilda objects revisited and updated}

\author[J. Correa-Otto et al.]{
J. Correa-Otto,$^{1}$\thanks{E-mail: jorge9895@gmail.com}
E. García-Migani$^{1}$
R. Gil-Hutton$^{1}$
\\
$^{1}$Grupo de Ciencias Planetarias, Dpto. de Geofísica y Astronomía, FCEFyN, UNSJ - CONICET, \\
 Av. J. I. de la Roza 590 oeste, J5402DCS Rivadavia, San Juan, Argentina\\
}

\date{Accepted XXX. Received YYY; in original form ZZZ}

\pubyear{2023}

\begin{document}
\label{firstpage}
\pagerange{\pageref{firstpage}--\pageref{lastpage}}
\maketitle

\begin{abstract}

In this paper, we perform a dynamical study of the population of objects in the unstable quasi-Hilda region. The aim of this work is to make an update of  the population of quasi-Hilda comets (QHCs) that have recently arrived from the Centaurs region. To achieve our goal, we have applied a dynamical criteria to constrain the unstable quasi-Hilda region that allowed us to select 828 potential candidates. The orbital data of the potential candidates was take from the ASTORB database and we apply backward integration to search by those that have recently arrived from the outer regions of the Solar System. Then we studied the dynamical evolution of the candidates from a statistical point of view by calculating the time-averaged distribution of a number of clones of each candidate as a function of aphelion and perihelion distances. We found that 47 objects could have been recently injected into the inner Solar System from the Centaur or transneptunian regions. These objects may have preserved volatile material and are candidates to exhibit cometary activity.

\end{abstract}




\section{Introduction}

In the outer edge of the main belt there is a dynamical group called Hilda asteroids. The limits of the Hilda region are 3.7 $\leq a \leq$ 4.2 au in semimajor axis, $e \leq$ 0.3 in eccentricity and $i \leq$ 20° in inclination. Objects in this region are near or trapped in the 3:2 mean motion resonance (MMR) with Jupiter \citep{schub68, schub82, schub91, nes97, ferraz98}. For the 3:2 MMR the resonant angle  librates around 0 degree \citep{zell85} and for small and medium amplitudes of libration there is a stable region in the resonance known as Hilda region where reside the stable asteroids of the group. However, close to the Hilda region we can find the quasi-Hilda comets \citep[QHCs,][]{kres79}, which could also be affected by the 3:2 MMR. The location of the QHCs is sometimes referred to as the quasi-Hilda region to indicate the dynamic difference between these objects and the Hilda asteroids. It is worth mentioning that a comet is an object characterized by its activity when it is close to the Sun.

The QHCs are Jupiter-family comets (JFCs) that have moved from outside to inside of Jupiter’s orbit \citep{kres79}. Moreover, the JFCs have evolved from the transneptunian region and have reached the zone of the Jupiter orbit after suffering the perturbations of the external planets during the period they behaved as Centaurs \citep{julio80, dun95, dun97}.

\begin{table}
\caption{List of QHCs identify by GG2016. A small tail of dust has been detected on the asterisked object \citep{gg18}.}             
\label{table:1}      
\centering                          
\begin{tabular}{c c  }        
\hline\hline  
Object  & Object     \\ 
\hline  
371837  & (450807)  2007 UC$_9$ \\
(18916) 2000 OG$_{44}$  & (457175)$^*$ 2008 GO$_{98}$ \\
 (507119)  2009 SR$_{143}$ & (524743) 2003 UR$_{267}$ \\
(577805) 2013 QR$_{90}$ & 2001 QG$_{288}$ \\
2002 UP$_{36}$ & 2006 XL$_5$  \\
 2009 KF$_{37}$  & \\
\hline                                  
\end{tabular}
\end{table}

However, QHCs are not the only objects inhabiting the quasi-Hilda region. There are also Hilda objects that have escaped from the stable Hilda region, which have a similar dynamic behaviour to JFCs \citep{disisto05}.As the Hilda region is located in the external zone of the main belt, where the asteroids are mainly D- and P-types, it is therefore not easy to physically distinguish them from the population of QHCs \citep{fit94, dal95, dal97, dal99, jew02, gb08}. Then, to distinguish comets from asteroids in the quasi-Hilda region, it is necessary to develop a dynamical study of the orbital evolution of these objects. The dynamical analysis of the quasi-Hilda region by \cite{tot06} found new members of this cometary group, and identified 23 objects that could be dormant or extinct cometary nuclei.

Recently, coma activity has been detected in the quasi-Hilda objects 2000 YN$_{30}$ \citep[presently known as 212P/NEAT,][]{chen13} and (457175) 2008 GO$_{98}$ \citep{leo17}, and numerical analysis indicate that in the past close encounters with Jupiter could have locked both bodies into a short-period orbit from a Centaur-like orbit. For this reason in \cite{gg16}  (GG2016 hereafter) we began a dynamical search for new QHCs candidates in the quasi-Hilda region. In that work we identify 11 candidates (see Table \ref{table:1}) that had a dynamical evolution showing they could have recently arrived from the outer Solar System. One of these objects, (457175) 2008 GO$_{98}$, was analyzed in \cite{gg18}, where we confirmed cometary activity.

In this work we follow the dynamical method described in GG2016 to update the population of quasi-Hilda comets that have recently arrived from the Centaur zone, which could become active near the perihelion of their orbits. In Sect. \ref{criterio}  we describe the method for the selection of the candidates and our results are presented and discussed in Sect. \ref{resultado}. Finally, our conclusions are summarized in Sect. \ref{conclusion}.

\section{Selection criteria} \label{criterio}

   \begin{figure*}
   \centering
   \includegraphics[width=1.9\columnwidth]{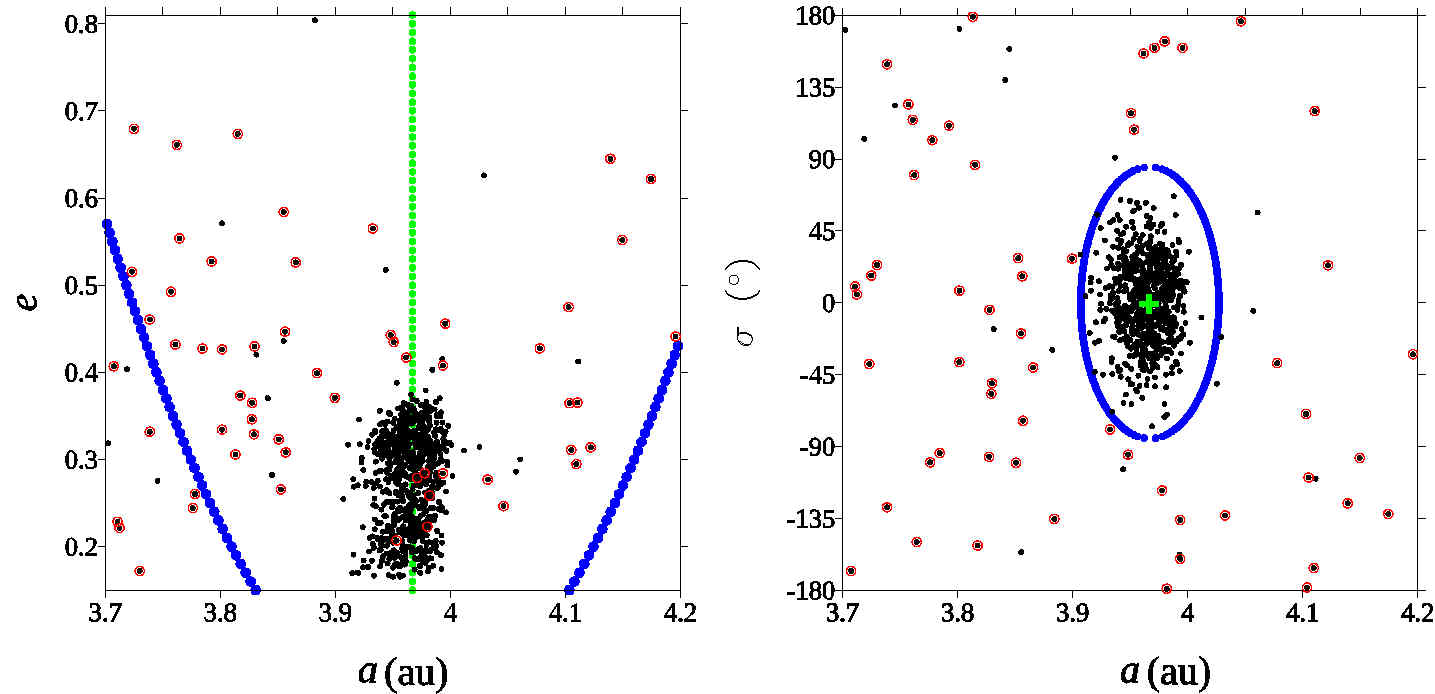}
   \caption{Left panel: Distribution of the pre-candidates in the ($a$, $e$)-plane, the nominal position of the resonance is indicated by a green line and the limits of the resonance are in blue line. Right panel: Distribution of the pre-candidates in the ($a$, $\sigma$)-plane, the center of the resonance is indicated by a green cross. The QHCs candidates are indicated by a red circle. We can see that the QHCs candidates are in the periphery of the MMR, beyond the blue ellipse in right panel.} 
              \label{Fig_dist}%
    \end{figure*}

From the ASTORB database\footnote{see ftp://ftp.lowell.edu/pub/elgb/astorb.html} we took as pre-candidates those objects of the quasi-Hilda zone with their osculating semimajor axis in the range 3.7 $\leq a \leq$ 4.2 au for the epoch 2458757.5 JD (Oct. 1, 2019), and an orbit determined by more than 180 days of orbital arc length. Then, following the same criteria as the one employed in GG2016 we found a final sample of 828 asteroids in the unstable quasi-Hilda zone.

These 828 pre-candidates were numerically integrated backward over time taking into account the perturbations of all the planets of the Solar System like in GG2016. The temporal evolution of each asteroid has been solved numerically by integrating the exact equations of motion using the same N-body integrator of GG2016, which uses the Bulirsh-Stoer code with a step size of 1 day and an adopted accuracy of $10^{-13}$. We found 47 new objects with a dynamical evolution that would indicate a recent income from the outer Solar System (see Table \ref{table:2}). This gives a total of 58 QHCs considering the 11 previously found in GG2016. An example of the possible evolution of these objects can be found in Fig. 1 of GG2016, where we can see the backward evolution of (18916) 2000 OG$_{44}$. The comet first achieve a Jupiter’s external orbit and then it reaches the transneptunian belt after being temporally captured by several mean-motion resonances.

In the left panel of Figure \ref{Fig_dist}  are show the 828 pre-candidates in the ($a$, $e$)-plane with the 58 QHCs candidates indicated by a red circle. We can see that almost all objects are close to the 3:2 MMR with Jupiter ($\sim$ 3.9607 au), whose nominal position is indicated by a vertical green line. Moreover, if we apply the simple pendulum model \citep[see][]{murray99} we can show the approximate position of the separatrix in blue line. We found that $\sim$ 98 \% of the objects are inside the resonance. It is worth to mention that there are more sophisticated models for the study of MMRs in the literature \citep[i.e.,][]{taba20, correa21}, however the pendulum model is the best option because we have only osculating elements of objects with a wide range of eccentricity and inclination values. So, the graphical representation that we present in the ($a$, $e$)-plane of Fig. \ref{Fig_dist} is not an exact dynamical picture, instead this is a simply way to show the influence of the resonance in the objects of interest.

\begin{table*}
\caption{QHCs candidates: orbital elements (2458757.5 JD) are in columns 2 to 4, absolute magnitude in column 5, time when the object reached $a>$ 5.2 au in column 6 and last column indicate the averaged time that the clones of each candidates stay in orbits with $q<$ 1 au.}             
\label{table:2}      
\centering                          
\begin{tabular}{c c c c c c c }        
\hline\hline  
Object  & $a$ & $e$ & $i$ & $H$ & $T$ & $<T>$     \\ 
        &  au &     &  $^\circ$ & & yr & yr \\
\hline 
7458 & 3.730 & 0.172 & 1.76 & 12.0 & -40913 & 24 \\
30512 & 3.850 & 0.323 & 25.70 & 12.8 & -18000 & 234 \\
85490 & 3.762 & 0.661 & 2.57 & 14.8 & -1021 & 589 \\
254010 & 3.982 & 0.258 & 17.05 & 13.9 & -11653 & 38 \\
424570 & 3.980 & 0.223 & 6.62 & 15.5 & -972 & 42 \\
431336 & 3.776 & 0.244 & 20.40 & 14.9 & -40212 & 59 \\
508861 & 3.815 & 0.673 &  4.16 & 17.5 & -8528 & 522 \\
551740 & 4.046 & 0.246 & 24.02 & 15.3 & -18091 & 149 \\
615767 & 4.122 & 0.314 &  9.86 & 15.9 & -20786 & 78 \\
2000 AC$_{229}$ & 4.149 & 0.551 & 52.43 & 16.8 & -30876 & 282 \\
2000 CA$_{13}$ & 3.764 & 0.553 & 1.45 & 18.0 & -5158 & 255 \\
2002 QD$_{151}$ & 3.710 & 0.228 & 4.75 & 15.4 & -925 & 55 \\
2004 QR$_{38}$ & 3.995 & 0.456 & 11.29 & 16.5 &  -8094 & 207 \\
2004 RP$_{111}$ & 4.174 & 0.622 & 14.14 & 18.2 & -11713 & 453 \\
2005 EC$_{272}$ & 3.738 & 0.460 & 7.15 & 16.9 & -5207 & 133 \\
2005 UK$_{380}$ & 3.899 & 0.370 & 3.03 & 17.8 & -23579 & 42 \\
2005 XR$_{132}$ & 3.761 & 0.431 & 14.47 & 16.4 & -12682 & 51 \\
2007 RM$_{150}$ & 3.856 & 0.446 & 11.11 & 16.1 & -18286 & 167 \\
2008 QZ$_{44}$  & 4.195 & 0.441 & 11.35 & 17.1 & -727 & 237 \\
2008 SZ$_{283}$ & 3.961 & 0.417 & 14.80 & 17.2 & -100 & 260 \\
2009 QM$_{24}$  & 3.784 & 0.427 & 13.75 & 16.4 & -1007 & 130 \\
2009 TC$_{54}$  & 3.817 & 0.373 &  5.87 & 16.4 & -1765 & 131 \\
2010 ES$_{189}$ & 4.103 & 0.364 & 31.61 & 15.6 & -2425 & 445 \\
2010 JB$_{184}$ & 3.977 & 0.284 & 21.18 & 16.4 & -2897 & 136 \\
2011 DL$_{12}$  & 3.829 & 0.329 & 13.48 & 16.7 & -1163 & 137 \\
2011 MX$_{9}$   & 4.103 & 0.475 & 19.23 & 16.7 & -15789 & 445 \\
2011 QQ$_{99}$  & 3.801 & 0.426 &  3.21 & 16.5 & -420  & 79 \\
2011 UB$_{301}$ & 3.801 & 0.334 & 21.13 & 17.0 & -10827 & 61 \\
2011 UG$_{104}$ & 3.993 & 0.407 & 29.59 & 16.2 & -4030 & 175 \\
2011 US$_{383}$ & 3.827 & 0.365 &  7.87 & 16.8 & -4088 & 22 \\
2011 WD$_{180}$ & 3.884 & 0.399 & 15.69 & 18.0 & -7447 & 146 \\
2014 MZ$_{101}$ & 4.110 & 0.365 & 17.62 & 16.3 & -1478 & 152 \\
2014 OM$_{449}$ & 3.993 & 0.284 & 19.81 & 18.0 & -31963 & 54 \\
2014 VF$_{40}$  & 3.757 & 0.492 & 24.48 & 16.3 & -4430 & 332 \\
2015 PV$_{306}$ & 4.109 & 0.294 & 12.41 & 15.3 & -900 & 103 \\
2016 CD$_{9}$   & 3.707 & 0.407 &  5.92 & 17.3 & -3060 & 52 \\
2016 JF$_{46}$  & 3.778 & 0.260 & 16.17 & 16.1 & -331 & 257 \\
2016 NQ$_{77}$  & 3.829 & 0.429 & 13.09 & 17.2 & -5797 & 74 \\
2016 WP$_{51}$  & 3.950 & 0.434 & 14.98 & 16.9 & -3348 & 92 \\
2016 BS$_{30}$  & 3.827 & 0.346 & 11.66 & 15.8 & -5646 & 36 \\
2017 FU$_{158}$ & 3.725 & 0.679 & 23.84 & 19.1 & -2373 & 560 \\
2018 PE$_{48}$  & 3.738 & 0.332 & 12.74 & 17.0 & -10630 & 17 \\
2019 JB$_{49}$  & 3.852 & 0.265 & 10.52 & 16.5 & -757 & 63 \\
2020 FV$_{35}$  & 3.813 & 0.305 &  4.12 & 17.0 & -1041 & 137 \\
2020 UO$_{43}$  & 4.139 & 0.645 &  1.75 & 18.7 & -565 & 563 \\
2020 XH$_{11}$  & 4.032 & 0.276 & 27.53 & 16.6 & -6247 & 599 \\
2021 JF$_{52}$  & 3.932 & 0.565 & 24.75 & 16.2 & -34527 & 359 \\
\hline                                  
\end{tabular}
\end{table*}

However, the semimajor axis of the objects do not indicate the real influence of the resonance. The reason of this is that the resonant action is at its mid-point (i.e., $\sim$ 3.9607 au) when the resonant angle is at its maximum amplitude and when the action is at its maximum, the angle is at its center of libration \citep[0° for the 3:2 MMR, see][]{zell85}. So, an object close to the critical semimajor axis could have a critical angle ($\sigma$) with a large amplitude, and therefore, it is not really close to the center of the MMR.

Hence, we can represent the distribution of the candidates in the ($a$, $\sigma$)-plane  by calculating the osculating value of the characteristic or resonant angle $\sigma$. Fig. \ref{Fig_dist} shows in the right panel the distribution of the 828 pre-candidates, with the QHCs candidates indicated by a red circle. We found here an interesting result: while the other quasi-Hildas objects can be placed anywhere in the plane, the 58 QHCs candidates are placed far from the resonant center. We can define a limit in semimajor axis of $\sim \pm$0.06 au from the central or nominal value ($a_0 \sim$ 3.9607 au) and a limit of $\sim \pm$90° in the resonant angle, and we can see that there are no QHCs candidates within the mentioned limits. This important result allows us to predict where we could find new QHCs candidates. Following \cite{tot06} we can define a region inside the unstable quasi-Hilda zone where the cometary objects can be found, and from the distribution of real objects we can propose the following empirical criteria for the limits in the ($a$, $\sigma$)-plane:

\begin{equation}
\begin{array}{c}
 \sigma = \pm 90 \sqrt{1- \left(  \dfrac{a-a_0}{0.06} \right)^2}  \,  \rm{,} \label{eq1}
\end{array}
\end{equation}

\noindent this ellipse is plotted in blue in the right panel of Fig. \ref{Fig_dist}. Then, the objects outside the ellipse are the possible QHCs candidates, and the objects inside the ellipse can be discarted.

\section{Results} \label{resultado}

   \begin{figure*}
   \centering
   \includegraphics[width=1.9\columnwidth]{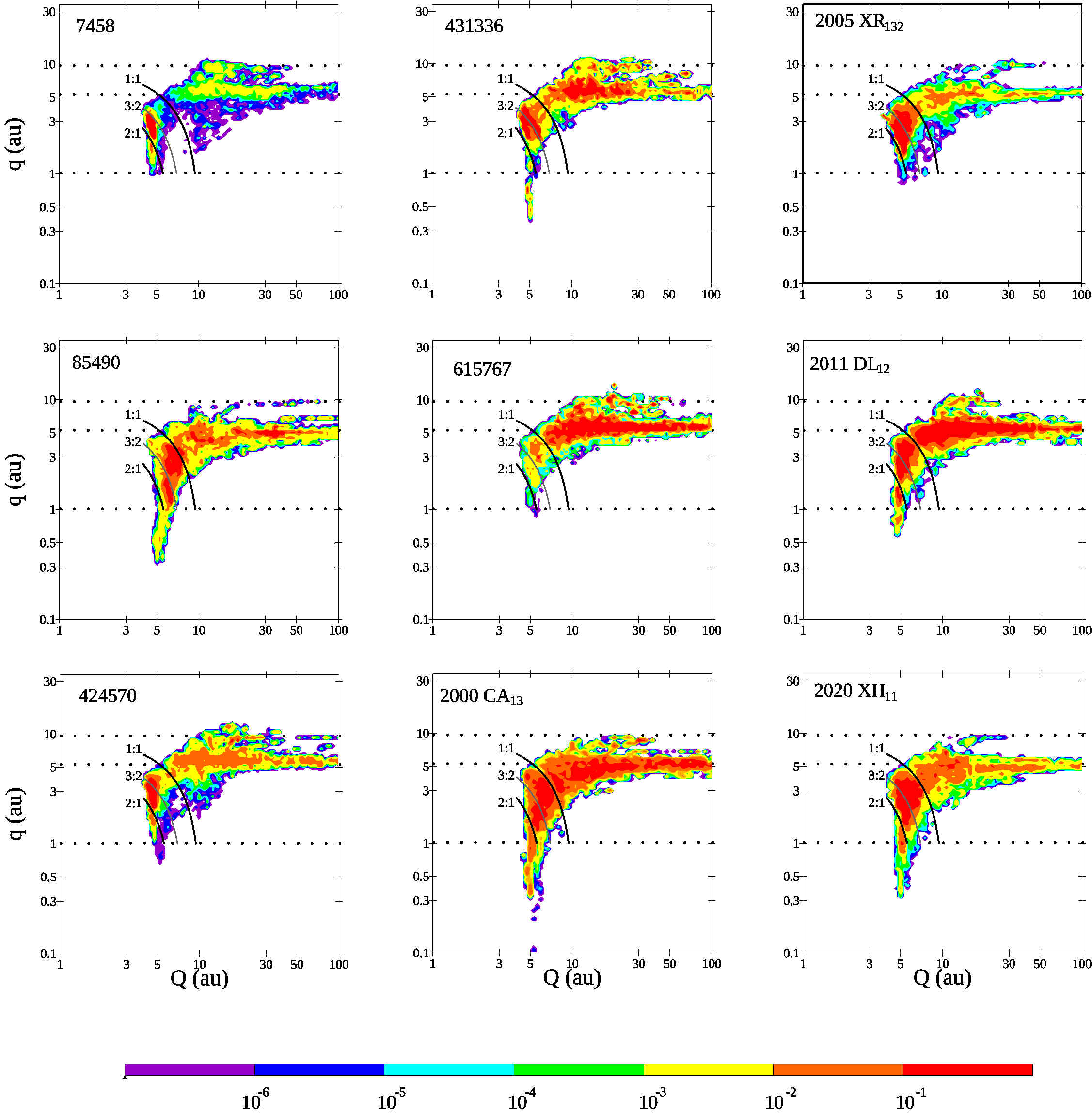}
   \caption{Probability distribution in the ($Q$, $q$)-plane for the clones of 9 candidates: 7458, 85490, 424570, 431336, 615767, 2000 CA$_{13}$, 2005 XR$_{132}$, 2011 DL$_{12}$ and 2020 XH$_{11}$. Dotted lines indicate perihelion distance of 9.5 au, 5.2 au and 1 au. Continuous lines indicate the 1:1 and 2:1 MMRs in blak and the 3:2 MMR in grey.} 
              \label{Fig_qq}%
    \end{figure*}

The recent dynamical history of the 58 objects of Tables \ref{table:1} and \ref{table:2} would indicate that they were Centaurs or transneptunian objects that evolve toward their actual position in the 3:2 MMR and ended as QHCs. However, the evolution of these objects is chaotic because they cross several MMRs and some of them have close approaches with Jupiter and other planets. Therefore, these chaotic orbits are sensitively dependent on initial conditions and the integration method employed (i.e., Burlish-Stoer or Radau). As in GG2016, the dynamical study of chaotic orbits could be developed from a statistical point of view by following the backward temporal evolution of clones of each object. These clones can be generated by small changes in the initial osculating orbital elements of each object like the initial conditions generated to calculate the maximum Lyapunov characteristic exponent \citep[LCE,][]{murray99}. 

   \begin{figure*}
   \centering
   \includegraphics[width=1.9\columnwidth]{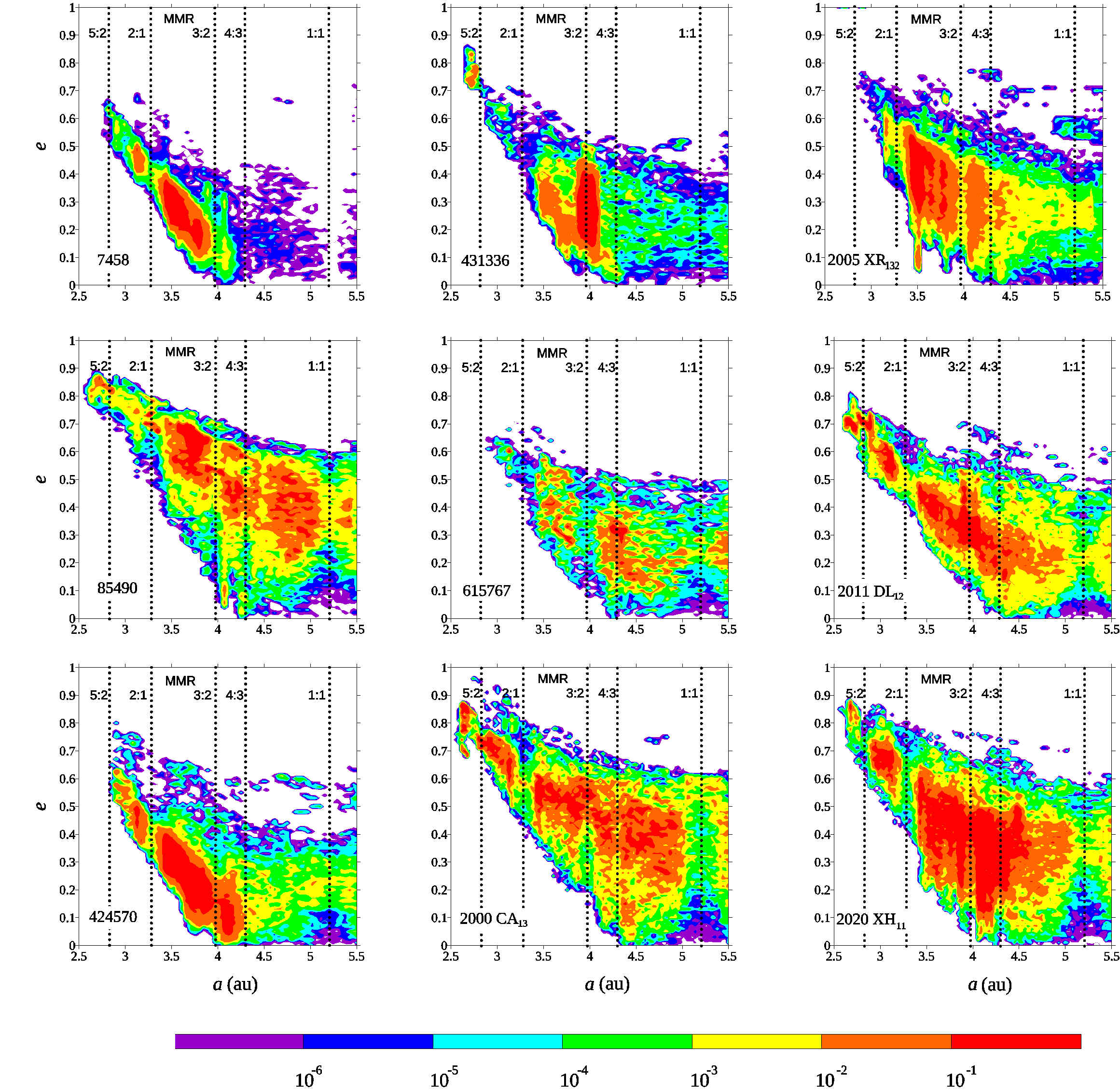}
   \caption{Probability distribution in the ($a$, $e$)-plane for the clones of 9 candidates: 7458, 85490, 424570, 431336, 615767, 2000 CA$_{13}$, 2005 XR$_{132}$, 2011 DL$_{12}$ and 2020 XH$_{11}$. Vertical dotted lines indicate the nominal position of the main MMRs with Jupiter.} 
              \label{Fig_ae}%
    \end{figure*}

As in GG2016 we study the  dynamical evolution of the 47 new QHCs candidates  by mapping the time-averaged distribution of 100 clones of each candidate in the plane of aphelion and perihelion distances, or the ($Q$, $q$)-plane \citep[e.g.,][]{tis03}. It is difficult to present the ($Q$, $q$)-planes corresponding to each candidate and, since we observe similar dynamical characteristics in all planes, we present the results of some QHCs candidates as examples.
Then, in Fig. \ref{Fig_qq} we show the results of 9 QHCs candidates as examples of the main dynamical characteristics observed after a backward integration of 50 kyr.

The results observed in the density maps of each one of the 47 QHCs candidates are similar and present the same characteristics obtained by GG2016 for their 11 candidates. We found a dynamical behavior consistent with objects coming from the outer Solar System (i.e., Centaur or transneptunian regions) that reach the quasi-Hilda region through the gravitational scattering of the giant planets. For the 9 candidates taken as examples in Fig. \ref{Fig_qq} we can see how Jupiter and Saturn produce a gravitational scattering over the clones, increasing their aphelion distances. These can be appreciated in the horizontal strips of density with almost constant perihelion distances at $q \sim$ 5.2 au and 9.5 au, which we have indicated by dotted horizontal lines. We can see an important density accumulation in the horizontal strip corresponding to Jupiter, and a lower density build-up in the strip corresponding to Saturn. This is due to the short time of integration of our simulations (50 kyr), the particles start closer to Jupiter and take more time to reach Saturn. Moreover, the timescales of particle ejection for Jupiter are shorter than that for Saturn. Numerical simulations with longer integration times would allow the particles to reach Neptune's orbit in a dynamical behavior similar to that founded in \cite{tis03} for the Centaurus population.

Moreover, almost all the 47 candidates studied are able to visit the region below 1 au where the comets could show intense activity. In Fig. \ref{Fig_qq} we indicate the limit $q=$ 1 au with a horizontal dotted line. As in GG2016 we calculate the averaged time that the clones of the QHCs candidates stay in orbits with a perihelion distance lower than 1 au ($<T>$), showing the result for each object in the last column of Table \ref{table:2}. As it is explained in GG2016, there is a high probability that a comet will become inactive if it remains in an orbit with $q<$ 1 au for at least 1 kyr. Then, QHCs candidates with a large average time in orbits with $q<$ 1 au have a high probability of being dormant or extinct cometary nuclei. Thus, that objects with $<T> $ less than $\sim$ 200 yrs are the most probable QHCs candidates that have recently arrived at the quasi-Hilda region and the best objects to search signals of activity due to outgassing. There are 28 candidates meeting this condition, which represent $\sim$60 \% of the sample of 47 QHCs candidates, a similar fraction to that obtained by GG2016 for their sample. Therefore, these objects represent an important opportunity to find more QHCs and to confirm the QHCs candidates.

On the other hand, as the clones start in a position close to the 3:2 MMR and then evolve through a process of slow diffusion we can see a bulk of density around this resonance. In Fig. \ref{Fig_qq} we indicate the nominal position of the 3:2 MMR by a gray line and we also include the nominal position of the 2:1 MMR ($a \sim$ 3.27 au) and the 1:1 MMR with black lines as reference. Then, to see in more details the dynamical characteristics of the evolution of these particles in the region $a<$ 5.2 au we represent the time-averaged distribution of the 100 clones of each candidate in the ($a$,$e$)-plane with limits $a \in (2.5,5.5)$ au and $e \in (0, 1)$. These results are present in Fig. \ref{Fig_ae}, where we also include the nominal position of the 5:2 ($\sim$ 2.82 au), 2:1 ($\sim$ 3.27 au), 3:2 ($\sim$ 3.96 au), 4:3 ($\sim$ 4.3 au) and 1:1 MMRs with Jupiter as vertical dotted lines. As the clones start in a position close to the 3:2 MMR we start with the analysis of this resonance. For some maps we can see a region of low density in the nominal position of the 3:2 MMR (i.e. 7458, 85490, 2005 XR$_{132}$), while other maps show a zone of high density (i.e. 431336, 2011 DL$_{12}$). This suggests that in some cases the 3:2 MMR is capable of transiently trapping clones. Instead, in all the cases we can see a region of low density around the 1:1 MMR and around the 2:1 MMR. These results suggest that the clones cross these resonances without ever being trapped, in contrast to the 3:2 MMR. Moreover, the dynamic evolution of the clones does not seem to be significantly affected by the 4:3 MMR. For this resonance, we neither see an accumulation that would indicate a temporary trapping nor a significant gap in the density. Finally, the clones reach the region of 3 au with a high eccentricity of $\sim$ 0.6 (i.e. $q \sim$ 1.2 au). Therefore, only the clones able to cross the 5:2 MMR ($\sim$ 2.82 au) could achieve the inner zone of the Solar System. However, it is worth noting that this resonance should not be understood as a barrier but as a reference to identify the candidates that can reach distances less than 1 au.

\section{Conclusions} \label{conclusion}

In this article we continue with the identification of QHCs candidates, initiated in GG2016. From the ASTORB database we have selected 828 pre-candidates to QHCs in the quasi-Hilda region, which were dynamically analyzed in order to find those that could have arrived from the Centaur region. The criteria used to select our sample were the same as those used in GG2016: i) $a \in (3.7, 4.2)$ au, ii) orbital arcs spanned by the observations greater than 180 days, and iii) the criteria of \cite{tot06} for unstable orbits apply to Lagrangian elements. Each pre-candidate was numerically integrated for a backward time-span of 50 kyr, considering the perturbation of all the planets in the Solar System and with the object assumed to be a massless body.

We report 47 QHCs candidates after our dynamical study. The backward integration of the orbit of each object showed a dynamical evolution from the quasi-Hilda region toward  the Centaur zone. This recent arrival of our candidates has taken place in a rather chaotic way due to the perturbation of the giant planets, so we have complemented the dynamical study of our candidates with a statistical analysis of the evolution of their orbits. Moreover, with our 47 candidates and the 11 reported by GG2016, we were able to define an empirical criterion in the resonance plane, which allows us to define a region where the QHCs candidates are most likely to be found.

From the results obtained we can deduce that all the QHCs candidates are able to visit the inner region of the Solar System and those with $<T>$ larger than 200 yr ($\sim$ 40 \% of the candidates) could be affected by strong activity and occasionally become inactive comets. Instead, the candidates whose clones remain below 1 au for a short time are more likely to still be active. Therefore, during the following 2–3 yr it is interesting to follow their orbits when they will pass by the perihelion because during that period they offer a good opportunity to detect activity.

\section*{Acknowledgements}

The authors gratefully acknowledge financial support by CONICET through PIP  112-202001-01227 and San Juan National University by a CICITCA grant for the period 2023-2024. The authors thank the referee, Serhii Borysenko, for his useful review, which led to an improvement of the paper. 

\section*{Data Availability}

Table \ref{table:2} indicates all the QHCs candidates and their main dynamical characteristics founded in our research. Moreover, the main results of the clones of some QHCs candidate are available in Fig. \ref{Fig_qq} and \ref{Fig_ae}. The obtained results for the other candidates and their clones will be shared on request to the corresponding author.



\bibliographystyle{mnras}
\bibliography{example} 

\bsp	
\label{lastpage}
\end{document}